\documentclass[12pt,preprint]{aastex}

\shorttitle{Currents in Solar Corona}
\shortauthors{S.R. Spangler}

\begin{document}

\title{A Technique for Measuring Electrical Currents in the Solar Corona}
\author{Steven R. Spangler}
\affil{Department of Physics and Astronomy, University of Iowa, Iowa City, IA 52242}

\begin{abstract}
A technique is described for measuring electrical currents in the solar corona. It uses radioastronomical polarization measurements of a spatially-extended radio source viewed through the corona.  The observations yield the difference in the Faraday rotation measure between two closely-spaced lines of sight through the corona, a measurement referred to as {\em differential Faraday rotation}.  It is shown that the expression for differential Faraday rotation is proportional to the path integral $\oint n \vec{B}\cdot \vec{ds}$ where $n$ is the plasma density and $\vec{B}$ is the coronal magnetic field. The integral is around a closed loop (Amperian Loop) in the corona. If the plasma density is assumed roughly constant, the differential Faraday rotation is proportional to the current within the loop, via Ampere's Law. The validity of the constant density approximation is discussed, and two test cases are presented in which the associated error in the inferred current is small, of order tens of percent or less. The method is illustrated with observations of the radio source 3C228 with the Very Large Array (VLA) in August, 2003.  A measurement of a differential Faraday rotation ``event'' on August 16, 2003, yields an estimate of $2.5 \times 10^9$ Amperes in the Amperian Loop.  A smaller event on August 18 yields an enclosed current of $2.3 \times 10^8$ Amperes.  The implications of these currents for coronal heating are briefly discussed.   
\end{abstract}


\section{Introduction}
Electrical currents certainly flow in the solar corona.  The structure seen in eclipse photographs or coronagraph images shows that pressure gradients must be balanced by electrodynamic forces.  Although information on the strength and form of the coronal magnetic field is limited, it is clear that the true field is deformed from the potential field generated by a magnetic scalar potential.  This difference is due to the presence of electrical currents.  In addition, there is a class of theories for heating of the solar corona which invoke Joule heating from coronal currents, probably contained in turbulent current sheets. This idea originated with \cite{Parker72}, and has been elaborated in many subsequent works \citep[see, for example, ][]{Gudiksen05, Peter06}.  

Observational measurements of coronal currents appear to be nonexistent.  In fact, measurements of the coronal magnetic field itself at heliocentric distances of a few solar radii are limited to results from Faraday rotation of a trans-coronal radio source and analysis of radio emission from solar flares \citep{Bird90}.  There is a literature on results from Faraday rotation observations of the large scale structure of the coronal field, as well as magnetic field inhomogeneities on a wide range of scales. Examples of such papers, which give references to the wider literature are \cite{Hollweg82,Bird90,Mancuso99,Mancuso00,Spangler05} and \cite{Ingleby07}. 

In this paper, I discuss how Faraday Rotation measurements can also provide an observational estimate of electrical currents in the corona.  The technique requires measurements of {\em differential Faraday Rotation} \citep{Spangler05}, which is the difference in the Faraday rotation measure between two closely-spaced lines of sight through the corona. As  discussed in the previously cited papers, a Faraday Rotation measurement yields the rotation measure, $RM$, given by \cite{Kraus66}
\begin{equation}
RM= \left( \frac{e^3}{8 \pi^2  c^3 \epsilon_0 m_e^2}\right) 
              \int_{LOS} n_e \vec{B} \cdot \vec{dz} 
\end{equation}
  The fundamental physical constants of $e, m_e, c, \mbox{ and } \epsilon_0$ are, respectively, the fundamental charge, the mass of an electron, the speed of light, and the permittivity of free space. The electron density in the plasma is $n_e$, and $\vec{B}$ is the vector magnetic field.  The incremental vector $\vec{dz}$ is a spatial increment along the line of sight, which is the path on which the radio waves propagate. Positive $dz$ is in the direction from the source to the observer. The subscript LOS on the integral indicates an integral along the line of sight. Equation (1) is in SI units, as opposed to cgs, which has been used in our previous papers.  The SI system is used in this paper for convenience in discussing electrical currents.  The units of the rotation measure are radians/m$^2$. 

As will be discussed in Section 2, when the rotation measures on two (or more) closely-spaced lines of sight are compared, one has an estimate of the electrical current between the two lines of sight.  Such multiple lines of sight are available when one images an extended radio source (such as a radio galaxy or quasar) which is occulted by the corona.  As is discussed in some of the papers referenced above, such measurements are straightforward with the Very Large Array radiotelescope of the National Radio Astronomy Observatory\footnote{The Very Large Array is an instrument of the National Radio Astronomy Observatory.  The NRAO is a facility of the National Science Foundation, operated under cooperative agreement with Associated Universities, Inc.}.

The present investigation was motivated by laboratory experiments using differential Faraday rotation measurements to determine internal currents in the Madison Symmetric Torus (MST) Reversed Field Pinch (RFP) at the University of Wisconsin \citep{Prager99} by D.L. Brower and W.X. Ding  \citep{Brower02,Ding03}.  In the case of the MST device, Ding and Brower use an infrared laser as the source of polarized radiation for measurement of Faraday Rotation.  Measurement of the rotation measure along spatially-separated paths is achieved by directing the beam along distinct chords through the machine. In spite of obvious technical differences, the physical situation is virtually identical to that of the present paper. The inferred current profiles in the MST, and their dependence on time, are in agreement with theoretical predictions and the results from other diagnostics of the plasma. The results of \cite{Brower02} and \cite{Ding03} encourage application of this technique to the solar corona.  

The organization of this paper is as follows.  Section 2 describes the basis of the technique, i.e. how differential Faraday Rotation measurements can provide a measurement of electrical currents in the corona.  Section 3 provides an observational implementation of this technique with Very Large Array observations of the radio source 3C228 in August, 2003 \citep{Spangler05}.  In that section, it is shown that the total current contained with the Amperian Loop formed by the lines of sight was as high as 2.5 GigaAmperes during one period on August 16, but  less than 0.8 GigaAmperes during a 3 hour period of high quality data before this event.  A smaller, marginal detection of differential Faraday Rotation on August 18 yielded a current  of the order of 0.23 GigaAmperes. In Section 4 we discuss the validity of an approximation used in the derivation of our expression for the current contained between two lines of sight through the corona.  This assumption (stated in eq.[3] below) is that the plasma density can be approximated as roughly constant in the region that makes the maximum contribution to the Faraday rotation, and taken outside the integral defining the differential rotation measure. Arguments based on a simple theoretical model for coronal current sheets as well as data from a laboratory fusion plasma indicate that the error introduced by this approximation is relatively small, of order tens of percent or less. Section 5 briefly discusses the implications of our observations and analysis for theories of coronal heating by Joule dissipation of the currents. Our estimate of Joule heating using the Spitzer resistivity is about 6 orders of magnitude less than the observational estimate for the volumetric heating rate in the relevant part of the corona. Finally, Section 6 summarizes and concludes.   

\section{Physical Basis of the Technique} 
The technique is based on simultaneous Faraday rotation measurements along two lines of sight through the corona, which are separated by a small angular distance on the sky, and a corresponding physical separation $l$ in the solar corona.  Such observations can and have been made with the Very Large Array (VLA) of radio galaxies and quasars \citep{Sakurai94,Mancuso99,Spangler05}.  A specific illustration is our observation of the radio galaxy 3C228 on August 16 and August 18, 2003 \citep{Spangler05,Spangler07}.  As may be seen in Figure 2 of \cite{Spangler05}, 3C228 is a double radio source with bright, highly polarized hot spots separated by about 46 arcseconds on the sky. The corresponding physical separation between the lines of sight to the two hot spots in the corona is about 33,000 km.  The observations of August 16 and August 18 were made when the lines of sight passed within $6.7 R_{\odot}$ and $5.2 R_{\odot}$ of the center of the Sun, respectively. 

A  Faraday Rotation measurement of an extended radio source with two components is illustrated in cartoon form in Figure 1(a).  The radio telescope measures rotation measure values $RM_A$ and $RM_B$ on the two paths which have a transverse separation $l$.  The shaded area is meant to represent the coronal plasma, with the gray scale conveying the strength and sign of the current density.  Black regions indicate regions of large positive current density, white areas are regions of large negative current density, and gray areas are regions of zero current density.  This picture is taken from numerical studies of current sheet development in MHD turbulence \citep{Spangler99}. 
\begin{figure}[h]
\epsscale{0.60}
\includegraphics[width=18pc]{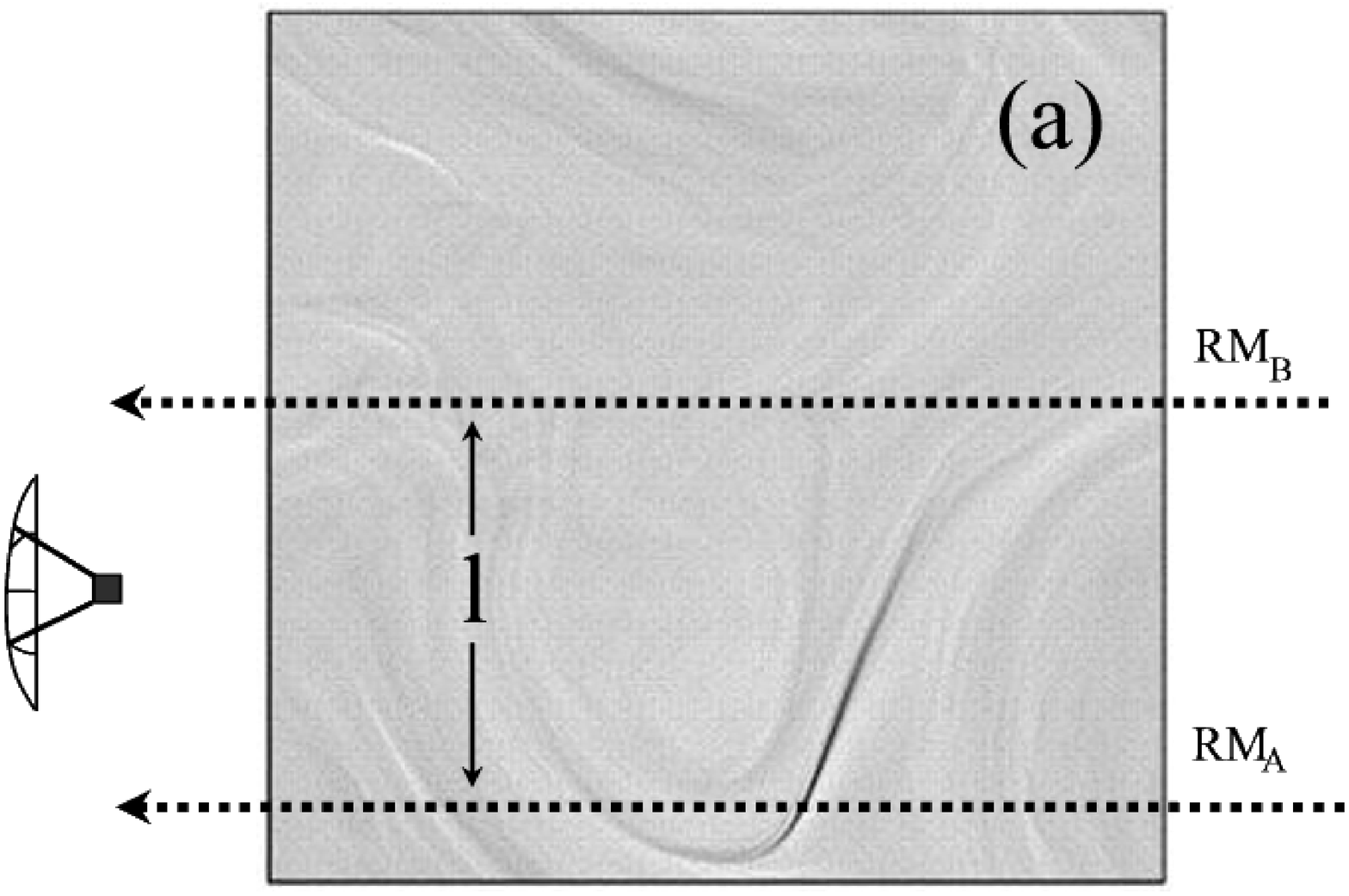}
\includegraphics[width=18pc]{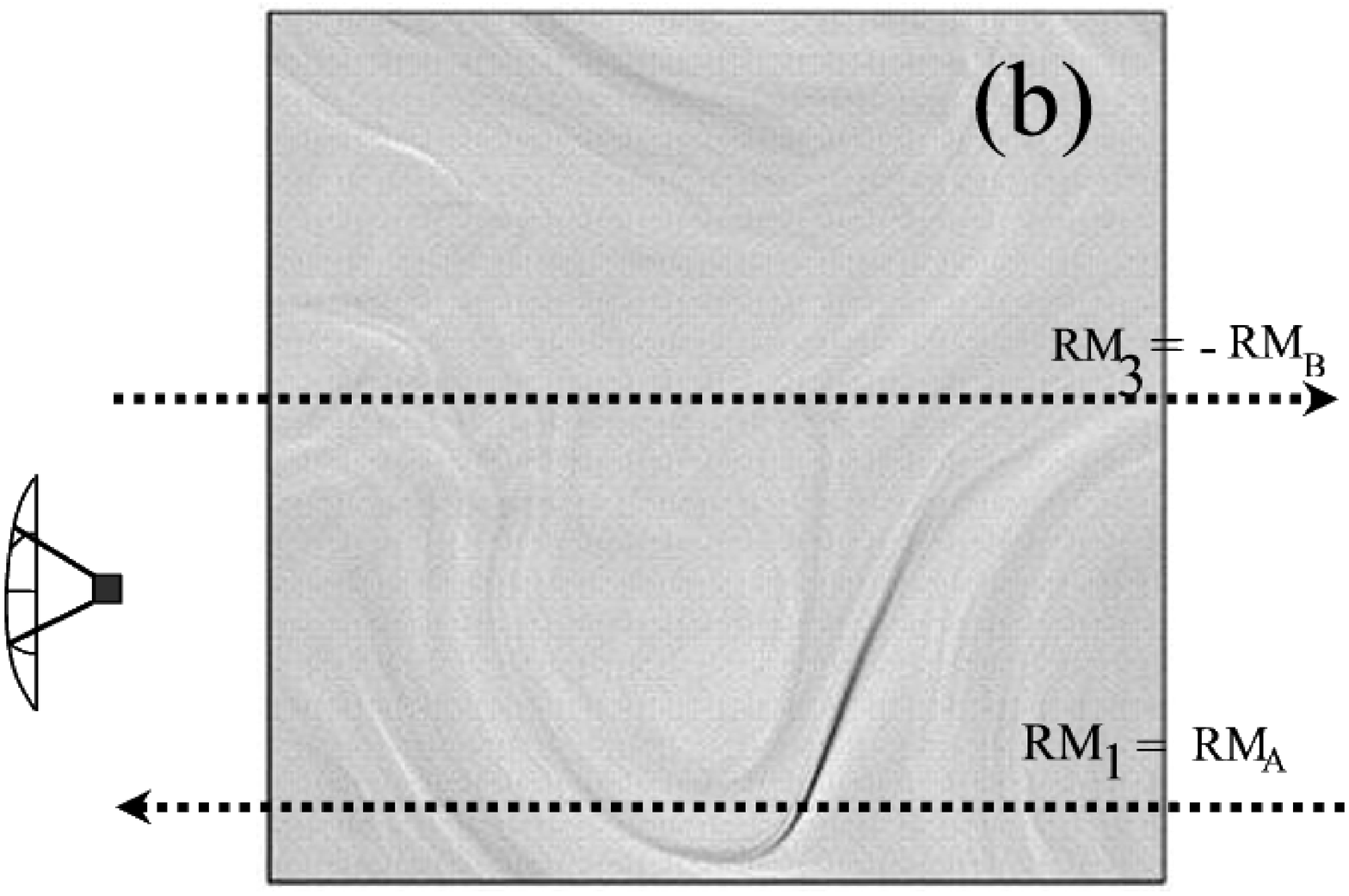}
\includegraphics[width=18pc]{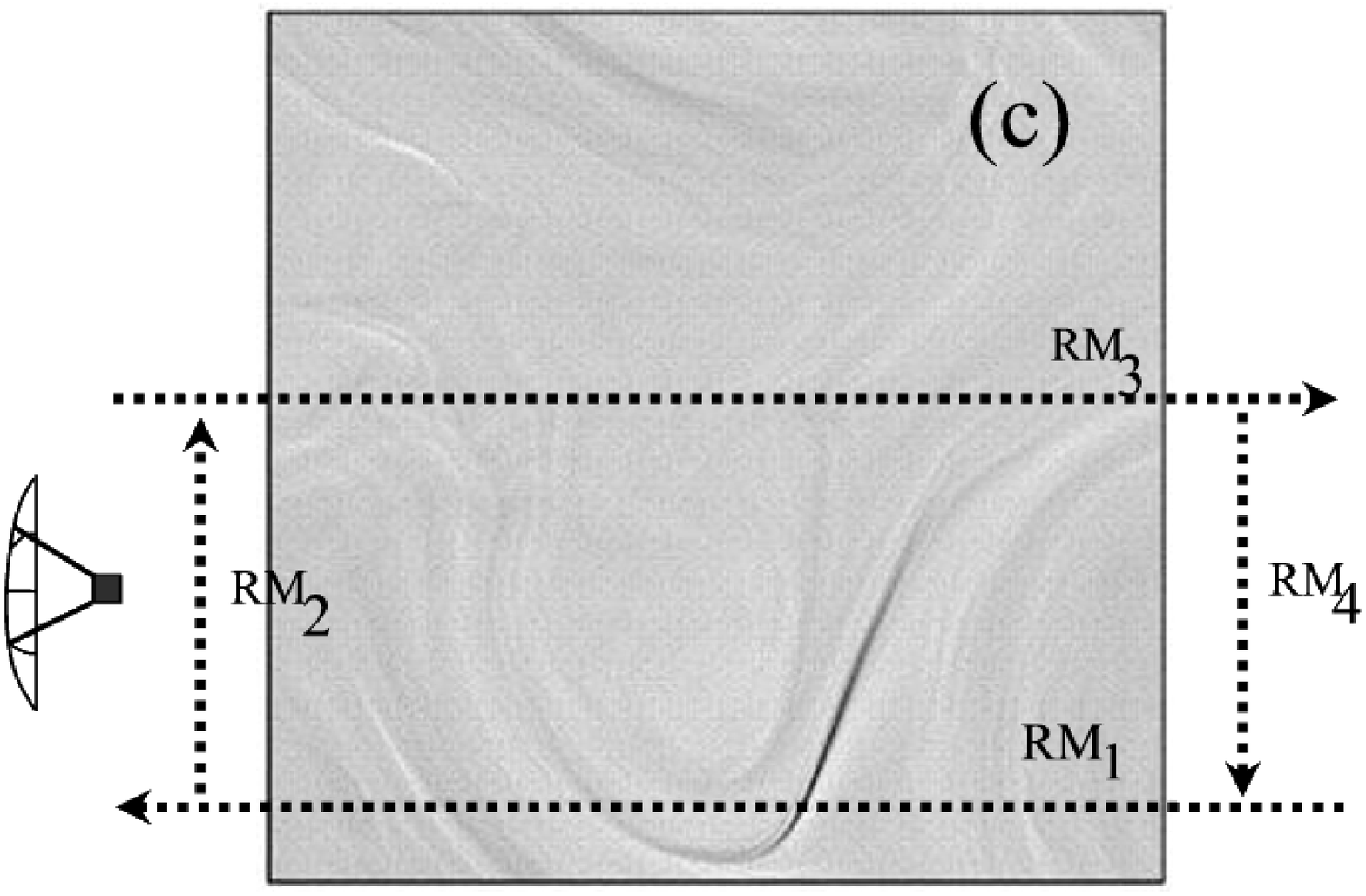}
\caption{Illustration of the measurement of coronal currents. Lines of sight through the corona are represented by dotted lines, with the arrow indicating the direction in which the integral is taken.  The square shaded area represents the coronal plasma, with the gray scale indicating the coronal current density. Positive current density is black, and negative is white.  (a) Illustrates the measurements of the Faraday rotation measure along two closely-spaced lines of sight, yielding $RM_A$ and $RM_B$. (b) Illustrates the differential rotation measure $\Delta RM \equiv RM_A - RM_B$.  It is equivalent to the sum of $RM_1$ and $RM_3$. (c) The differential Faraday rotation measurement is very nearly the same as the sum of the four parts $RM_1+RM_2+RM_3+RM_4$, since the ``end pieces'' $RM_2$ and $RM_4$ contribute negligibly to the sum.  The near-equality of the sum of the four segments and the differential Faraday Rotation measurement means that $\Delta RM$ is essentially equal to the path integral of $n\vec{B}$ around the Amperian Loop shown in panel (c).  The gray scale representation of the coronal plasma is taken from Figure 6 of \cite{Spangler99}. }
\label{cartoon}
\end{figure}

The distribution of currents shown in Figure 1 is also consistent with a picture in which the current-carrying entities are coronal loops. This case would be relevant to streamer plasmas and the closed-field part of the corona.  In this case, currents flow along coronal loops, up on one footpoint and down on the other, as discussed, for example, by \cite{Chen89} and \cite{Chen07}. The lines of sight to a radio source in a differential Faraday rotation experiment could ``bracket'' one branch of a loop and miss the other, so the cartoon in Figure 1 reasonably represents this case as well.   

The differential Faraday Rotation is defined as $\Delta RM = RM_A - RM_B$.  As is illustrated in Figure 1(b), $\Delta RM$ is equivalent to the sum $RM_1 + RM_3$, where $RM_1$ is the same as $RM_A$, and $RM_3$ is the same as $RM_B$, but with the direction in which the integration is taken being reversed, i.e. from the telescope to the source.  The value of $\Delta RM$ differs insignificantly from the sum of four terms, $RM_1 + RM_2 + RM_3 + RM_4$, as shown in Figure 1(c).  The reason for this is that the contributions of terms $RM_2$ and $RM_4$ are small compared to that of $RM_1 + RM_3$.  There are two reasons for this.  First, for the circumstances of a real coronal observation, the length of the segments 2 and 4 is small compared to 1 and 3.  Second, these segments may be considered to be located far from the point of closest approach to the Sun (taken to be the middle of the diagram in all three panels), where both the plasma density and magnetic field strength are much smaller than their values near the point of closest approach.  

The net result then, is that the differential Faraday Rotation is given by 
\begin{equation}
\Delta RM = RM_1 + RM_2 + RM_2 + RM_3 = C \oint n_e \vec{B} \cdot \vec{ds}
\end{equation}
where the last expression represents a path integral around the closed Amperian Loop defined by the segments 1,2,3, and 4 in Figure 1.  The constant C is defined as $C=\frac{e^3}{8 \pi^2 c^3 \epsilon_0 m_e^2} = 2.631 \times 10^{-13}$ in SI units. 

Equation (2) strongly recalls Ampere's Law, but the obvious difference is that in the present case the integrand in (2) is not just the magnetic field, but the product of the plasma density and the magnetic field.  If we make the assumption that the measured $\Delta RM$ is dominated by a region in which the plasma density is relatively uniform, and given by a value $\bar{n}$, we then have 
\begin{equation}
\Delta RM = C \oint n_e \vec{B} \cdot \vec{ds} \simeq C \bar{n}\oint \vec{B} \cdot \vec{ds}
\end{equation}
and Ampere's Law can be utilized.  The remainder of this paper will assume the convenience of equation (3).  Although we think that equation (3) is a reasonable first-order approximation to employ, it is legitimate to ask for an estimate of the associated error, at least in some plausible test cases.  This matter is discussed in Section 4 below. 

Given the approximation of equation (3), we can now use Ampere's Law, 
\begin{equation}
\oint \vec{B} \cdot \vec{ds} = \mu_0 I
\end{equation}
where $\mu_0$ is the permeability of free space and $I$ is the current contained within the Amperian Loop.  Use of equation (4) in equation (3) allows us to write an equation giving the coronal current in terms of the measured $\Delta RM$, 
\begin{equation}
I = \frac{\Delta RM}{C \mu_0 \bar{n}}
\end{equation}

Use of equation (5) requires an estimate of $\bar{n}$.  Fortunately, there are a number of empirical expressions for the coronal plasma density as a function of heliocentric distance.  For the purposes of this paper, we utilize the following expression \citep{Spangler05}, which has been employed in previous analyses of Faraday Rotation observations, and is in good agreement with independent estimates. 
\begin{equation}
n(r) = 1.83 \times 10^{12} \left( \frac{r}{R_{\odot}} \right)^{-2.36} \mbox{ m}^{-3}
\end{equation}
 For purposes of simplifying the subsequent formulas, the index 2.36 will be rounded off to 2.5.  

The density given by equation (6) would be the maximum that would be measured along a line of sight with a dimensionless impact parameter $R_0 = \frac{r_{min}}{R_{\odot}}$, where $r_{min}$ is the smallest heliocentric distance along the line of sight.  The density $\bar{n}$ in equation (5) represents an average over the Amperian Loop defined by the observations.  Therefore, for a given impact parameter $R_0$, $\bar{n}$ should be less that $n(R_0)$ by some factor which is defined by $\bar{n}=\alpha n(R_0)$.  With these parameterizations, substitution of (6) into (5) yields the formula that will be used in Section 3, 
\begin{equation}
I_{obs} = 1.65 \times 10^6 \frac{(\Delta RM) R_0^{2.5}}{\alpha} \mbox{   Amperes}
\end{equation}
The subscript ``obs'' on the current indicates that it is an estimate computed from observable quantities.  
Before leaving this section, it is worthwhile to emphasize the fundamental fact  that the current given by equation (7) is the sum of all currents, positive and negative, within the Amperian Loop.  It is thus obviously possible for strong currents to be flowing, but the total current given by equation (5) or (7) to be zero. 

The approach described above requires that the region which makes the dominant contribution to the Faraday rotation measure be sufficiently uniform to allow definition of a meaningful mean density $\bar{n}$. As will be seen in Section 4.2 below, modest variations in plasma density do not result in a significant quantitative error when equations (5) or (7) are applied.  However, larger errors could occur if one line of sight passes through a physically distinct region relative to the other.  Whether this situation is pertinent will be determined by future observations which apply the technique described in this paper, and benefit from independent diagnostics of the plasma along the line of sight.  

\section{Observational Implementation and Estimates of Coronal Currents}
We use data from the VLA observations of 3C228 in August, 2003.  Further details of the observations and data analysis are given in \cite{Spangler05} and \cite{Spangler07}. The characteristics of the observations relevant to the present discussion are given at the beginning of Section 2 of the present paper.  

Figure 3 of \cite{Spangler05} shows the Faraday Rotation \footnote{The rotation in the polarization position angle, which is the quantity measured. It is given by the product of the rotation measure and the square of the observing wavelength.} time series for 3C228 on August 16, 2003.  The main feature of those observations was a large change in rotation measure over the 8 hour duration of the observing session.  The data shown there distinguished measurements of the Faraday rotation to three source components, the two hot spots and a portion of the radio jet which lies between the hot spots \citep[see Figure 2 of ][]{Spangler05}.  Similar data are available for observations of 3C228 on August 18 \citep{Spangler07}.

These data have been used to calculate time series of $\Delta RM$ for both observing sessions.  These data are shown in Figure 2, and are the differences in the rotation measures to the two hot spots of 3C228. The values of $\Delta RM$ plotted have used the data at both frequencies of observation, 1465 and 1665 MHz.  In the case of the observations of August 18, 2003, only data from the last 4 hours of the observing session are used.  Prior to this time, elevated system temperatures due to the proximity of the Sun caused a substantial loss in data quality.  
\begin{figure}[h]
\epsscale{0.70}
\includegraphics[angle=-90,width=20pc]{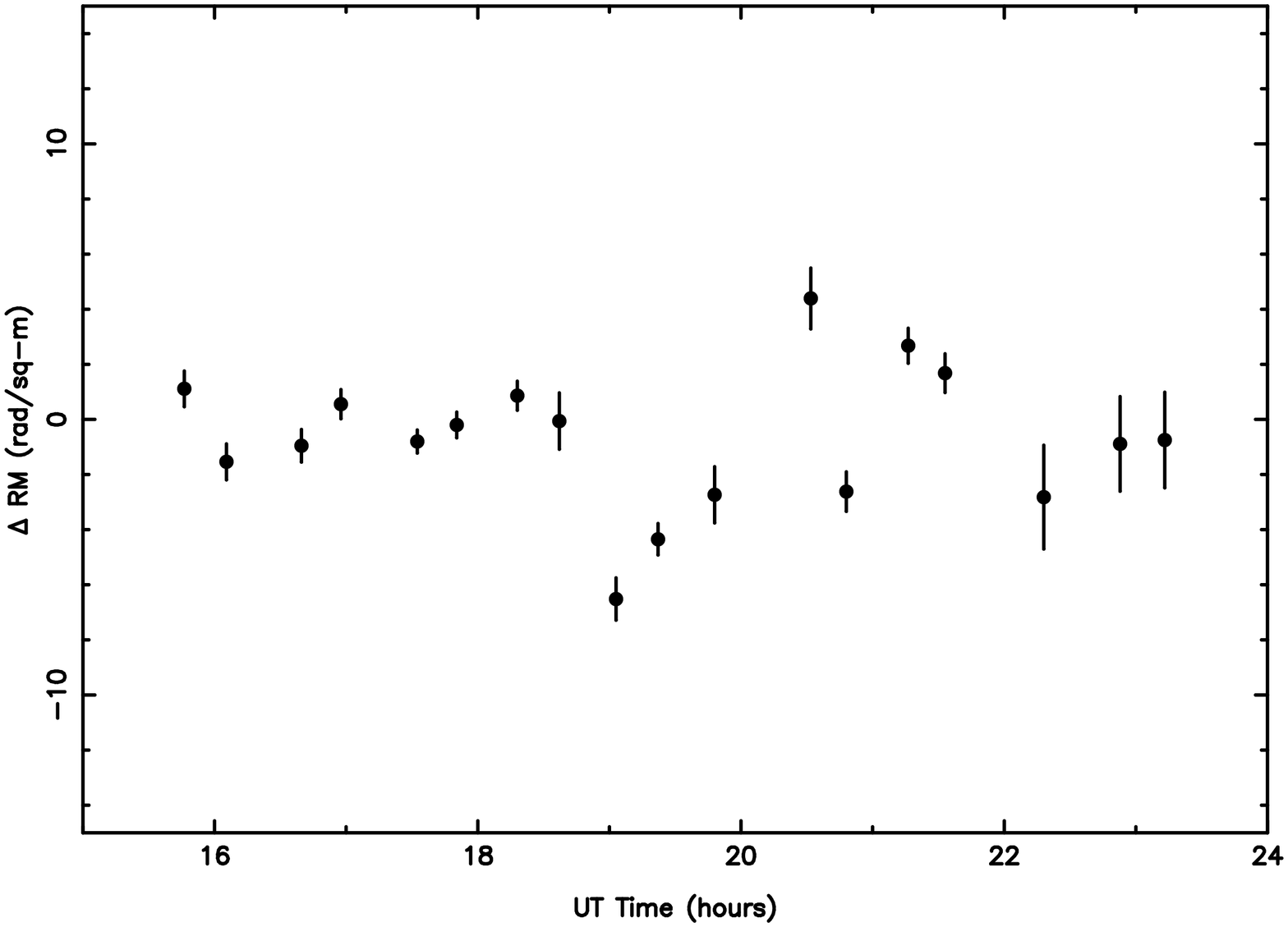}
\includegraphics[angle=-90,width=20pc]{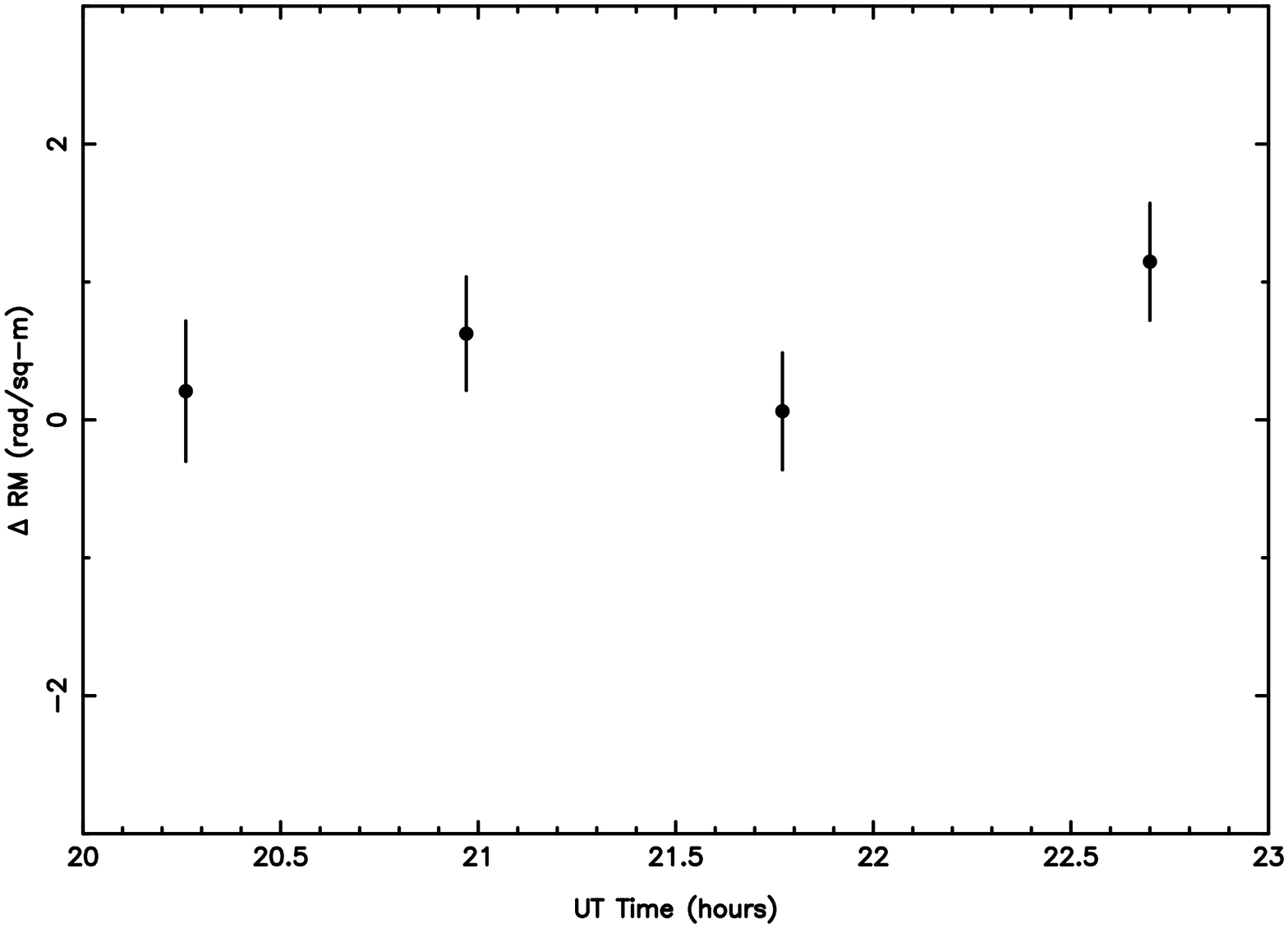}
\caption{Differential Faraday Rotation data for August 16, 2003 (left panel) and August 18, 2003 (right panel).  The measurements show the difference in the rotation measures to the hot spots of the radio source 3C228, which are separated by 46 arcseconds (33,000 km in the corona). The smaller number of measurements on August 18 results from reduced data quality prior to 20h UT, due to solar interference.  }
\label{data}
\end{figure}
The most prominent signal in the data for either day is the ``event'' at 19h UT on August 16.  This was discussed in \cite{Spangler05} and evidence for its credibility is given in \cite{Spangler07}.  The differential Faraday Rotation had a maximum absolute value of $6.52 \pm 0.77$ rad/m$^2$ in the scan at 19h UT, and declined over the next two scans to a value around zero.  

The observations for August 18, when the line of sight passed closer to the Sun, show no such prominent event.  The final scan shows a marginally  significant detection of $\Delta RM = 1.15 \pm 0.42$ rad/m$^2$.  Other measurements on this day are consistent with zero.  

Returning to the data for August 16, the three hours prior to the 19h event show measurements consistent with $\Delta RM = 0 \pm 2$ radian/m$^2$.  These data can be used to set a significant upper limit to the current within the lines of sight.  Data after 20h UT on August 16 appear to show significant $\Delta RM$ values, although with no indication of a systematic mean value.  We do not believe these data show good evidence for differential Faraday Rotation during this interval, because the errors are calculated from a propagation of known radiometer noise errors, and may (and probably do) underestimate the true errors due to solar interference.  

The final observational point to be made is that the $\Delta RM$ event at 19h on August 16 occurred at a time when the rate of change of the rotation measure increased \citep[see Figure 3 of ][]{Spangler05}.  As mentioned in that paper, the line of sight during this time was moving deeper into a coronal streamer, so this feature may not be typical of coronal properties. In particular, it may reveal the system of currents associated with a streamer rather than a network of turbulent current sheets as illustrated in Figure 1, and which constitutes the model in the calculations to be briefly described in Section 5.   

The above data can be used in equation (7) to obtain estimates for the electrical current.  The results are given in Table 1, in which results are given for 3 periods during August 16 and 18, 2003. In the calculations used here, we have assumed a value of $\alpha$ (defined above, just prior to equation (7)) of $\frac{1}{2}$. In Table 2 we use the absolute value of $\Delta RM$, since its sign, and the corresponding sign of $I$, are not considered in this paper.   

\clearpage
\begin{deluxetable}{crrr}
\tabletypesize{\small}
\tablecaption{Results on Coronal Currents\label{tbl-1}}
\tablewidth{0pt}
\tablehead{\colhead{Date} & \colhead{UT Time (hours)} & \colhead{$\Delta RM$ (rad/m$^2$)} & \colhead{$I$ (Amperes)} }
\startdata
Aug. 16& 15.5 - 18.5 & $\leq 2.0$ & $ \leq 7.7 \times 10^8$ \\
Aug. 16& 19.0 & $6.52 \pm 0.77$ & $(2.50 \pm 0.30) \times 10^9$ \\
Aug. 18& 22.7 & $1.15 \pm 0.42$ & $(2.34 \pm 0.86) \times 10^8$ \\
\enddata
\end{deluxetable}
From Table 1 it can be seen that detectable differential Faraday rotation requires currents of $10^8 - 10^9$ Amperes.  Upper limits deduced from present radioastronomical observations will then be in this range as well.  

Before leaving this section, we again emphasize that differential rotation measures $\Delta RM$ of even a few radians/m$^2$ are rare in our two August, 2003 observing sessions spanning a total observing time of 16 hours.  The apparent detection of coronal currents on August 16 was an isolated event, and may be associated with a coronal streamer.  The event on August 18 is of marginal statistical significance.  Consequently, the typical magnitude of differential Faraday rotation through the corona at the heliocentric distances characteristic of these observations is probably less than that of the event on August 16, with net currents which are also correspondingly smaller.    

\section{Validity of the Constant Density Approximation}
In this section we discuss the validity of the approximation employed in equation (3). The motiviation in adopting it is that it is a reasonable, first order approximation which permits application of Faraday's Law. The discussion in this section is organized into four topics. 
\subsection{The Nature of Coronal Current Sheets} It is obvious that until we have a better understanding of the physical nature of coronal current sheets, including their compressibility and structure, we will be unable to determine how close (3) is to an equality. The most obvious counterexample to (3), that in which the plasma density varies in a constant magnetic field, is not a satisfactory model for coronal plasma structures. In general, a spatial gradient in the plasma density results in a similarly directed gradient in the pressure.  If the plasma structures are in approximate mechanical equilibrium, this pressure gradient must be balanced, and the most obvious possibility is a $\vec{J} \times \vec{B}$ force, where $\vec{J}$ is the current density.  Thus currents will be necessary where there are plasma density variations, and while  the two sides of equation (3) might not be equal, they should be of comparable magnitude.

A case in which the above considerations are not valid would be if differential Faraday rotation signals are due to parallel-propagating slow magnetosonic waves, for which there are density variations but no field variations. There is also the possibility that density variations are associated with anticorrelated temperature variations, so that pressure variations are absent, and the need for currents to achieve mechanical equilibrium is removed.  

\subsection{The Z-Pinch as a Model for Coronal Current Filaments}
In this subsection we adopt a simplified model for coronal current filaments to explicitly calculate the error associated with the approximation (3). The model we adopt is the Z-Pinch \citep{Krall73,Gurnett05} which is a solution to the equation for mechanical equilibrium
\begin{equation}
\vec{J}\times\vec{B} - \nabla p = 0
\end{equation}
in cylindrical symmetry, where $p$ is the plasma pressure. In the Z-Pinch, current can flow in the z direction (coordinate perpendicular to the plane of the page in Figure 1,  not the coordinate of the observational line of sight defined in eq.[1]) and $\theta$ direction (azimuthal direction in the plane of Figure 1), and thus balance pressure gradients in the radial direction in this plane.  The Z-pinch thus resembles the current sheets shown in Figure 1, with the exception that we now approximate these current filaments as having circular symmetry.  

Although the Z-Pinch is an unstable equilibrium \citep{Krall73,Gurnett05}, it is a well-studied model for current-carrying plasma structures, and one can imagine compressible astrophysical turbulence as consisting of such structures forming, persisting for a while, then dispersing with subsequent reformation. 

We adopt the common simplification of ignoring the $\theta$ component of the current, which is equivalent to assuming no radial gradients in the $z$ component of the magnetic field. Equation (8) then yields the following differential equation for the relation between the plasma pressure and the $\theta$ component of the magnetic field
\begin{equation}
\frac{B_{\theta}}{r}\frac{d(r B_{\theta})}{dr} = -\mu_0 \frac{dp}{dr}
\end{equation} 
where $r$ in this section is the radial coordinate in a cylindrical coordinate system, not the heliocentric distance as used in Section 2. It is $B_{\theta}$ that causes the differential Faraday rotation in this model. In discussions of the Z-Pinch  \citep{Krall73,Gurnett05}, one prescribes one of the fields $p$ or $B_{\theta}$, then uses equation (9) to determine the other.  This procedure is well-suited to our present task, which is to see how spatial gradients in the plasma density of a plausible plasma structure affect differential Faraday rotation.  

We assume an isothermal approximation for the plasma equation of state, which will give a larger density change for a given pressure change than adiabatic equations of state. 
\begin{equation}
p(r) = c_s^2 \rho = c_s^2 m_H n(r)
\end{equation} 
where $c_s$ is the sound speed, $\rho$ is the mass density, $m_H$ is the mass of a hydrogen atom, and $n(r)$ is the number density of atoms.  We assume a pure hydrogen plasma.  Substituting equation (10) into (9), and making the change of variable $B_{\theta} \rightarrow rB_{\theta} \equiv f$, equation (9) becomes
\begin{equation}
\frac{df^2(r)}{dr} = -2 \gamma r^2 \frac{dn(r)}{dr}
\end{equation}
with $\gamma \equiv \mu_0 c_s^2 m_H$, which is directly integrable.  The solution is 
\begin{equation}
f^2(r) = (rB_{\theta}(r))^2 = -2 \gamma \int_0^r r^2 \frac{dn}{dr}dr
\end{equation} 

We adopt a model for the density profile which is enhanced in the current channel, 
\begin{equation}
n(r) = \bar{n} + n_1e^{-(r/a)}
\end{equation}
 with $a$ being the specified radial extent of the current filament. We substitute equation (13) into (12), and change variable of integration from $r \rightarrow \xi \equiv \frac{r}{a}$.  The solution then becomes 
\begin{eqnarray}
f^2(r) = (rB_{\theta}(r))^2 = 2 \gamma n_1 a^2 F(\xi)\\
F(\xi) \equiv \int_0^{\xi} \xi^2 e^{-\xi}d\xi = 2-(\xi^2 + 2\xi +2)e^{-\xi}\\
B_{\theta}(r) = \frac{\sqrt{2 \mu_0}}{r} \sqrt{\Delta p} a \sqrt{F(\xi)}
\end{eqnarray}
where $\Delta p \equiv c_s^2 m_H n_1$.
Equation (16) also yields the expression for the current density 
\begin{eqnarray}
J_z(r)=\left( \frac{1}{a} \right)\sqrt\frac{\Delta p}{2 \mu_0} G(\xi)\\
G(\xi) \equiv \frac{\xi e^{-\xi}}{\sqrt{F(\xi)}}
\end{eqnarray} 

We now express the different fields in the current filament in terms of common parameters which characterize the host plasma and the current channel.  The plasma pressure is, from above, 
\begin{equation}
p = c_s^2 m_H (\bar{n} + n_1 e^{-\xi})
\end{equation}
We define the ``modulation index'' $\epsilon \equiv \frac{n_1}{\bar{n}} = \frac{\Delta p}{p_0}$, so that 
\begin{eqnarray}
n=\bar{n}(1 + \epsilon e^{-\xi})\\
p=p_0(1 + \epsilon e^{-\xi})
\end{eqnarray}
With $p_0 \equiv c_s^2 m_H \bar{n}$. We now relate the plasma pressure to the magnetic pressure via
\begin{eqnarray}
p_0=\beta (\frac{B_0^2}{2 \mu_0})\\
\Delta p = \epsilon \beta (\frac{B_0^2}{2 \mu_0})
\end{eqnarray}
Where $\beta$ is the conventional plasma $\beta$, here defined as the ratio of gas pressure to magnetic pressure, and $B_0$ is the mean magnetic field.  
Using equation (23) in the equation for $B_{\theta}$ gives
\begin{eqnarray}
B_{\theta} = B_0 \sqrt{\epsilon \beta} F_2(\xi) \\
F_2(\xi) \equiv \frac{\sqrt{F(\xi)}}{\xi}
\end{eqnarray}

With these expressions, we can now calculate both sides of the near-equality (3).  The integral is taken in a clockwise sense, as indicated in Figure 1(b), on the Amperian Loop shown in Figure 1, with the convention that the origin of the polar coordinate system is midway between the integration paths A and B. The results of the calculation below are not qualitatively changed if the center of the current filament ($r=0$) is not midway between the two linear paths of integration A and B.  The segments of the Amperian Loop connecting paths A and B (segments 2 and 4 in Figure 1(c)) are assumed to lie at infinity and are neglected in the calculation below. 

The integral around the Amperian Loop is then given by symmetry considerations as  
\begin{equation}
\oint n \vec{B} \cdot \vec{ds} = 4 \int_b n \vec{B} \cdot \vec{ds}
\end{equation}
where the path segment b corresponds to  $\frac{\pi}{2} > \theta > 0$.
In what follows we evaluate the integral over path b.  Given the expression for $B_{\theta}$, etc developed above, the integrand of equation (3) is 
\begin{equation}
n \vec{B} \cdot \vec{ds} = n(\xi) B_{\theta}(\xi) \hat{e}_{\theta} \cdot \hat{e}_x dx
\end{equation}
where $\hat{e}_x$ is a unit vector in the the x direction (horizontal axis in Figure 1), and $\hat{e}_{\theta}$ is the unit vector in the direction of increasing $\theta$.  The dot product of the unit vectors is $\hat{e}_{\theta} \cdot \hat{e}_x = -\cos \bar{\theta}$, where $\bar{\theta}$ is the complement of $\theta$.  

An important variable is the perpendicular distance $R$ of the integration path from the center of the current channel. In terms of our observations, $2R$ would be the separation $l$ between the lines of sight through the corona.  The incremental path element $dx = R \sec^2 \bar{\theta} d \bar{\theta}$, and the variable $\xi$  is $\xi=\frac{R}{a} \sec \bar{\theta}$. Substituting all of these relations into (27) and forming the integral, we have 
\begin{equation}
\oint n \vec{B} \cdot \vec{ds} = 4 \int_0^{\pi/2} n(\xi) B_{\theta}(\xi) R sec \bar{\theta} d\bar{\theta} 
\end{equation}
 Substituting equations (20) and (24) into (28), we have
\begin{eqnarray}
\oint n \vec{B} \cdot \vec{ds} = 4n_0B_0 \sqrt{\epsilon \beta}R 
\left[\int_0^{\pi/2} F_2(\xi) \sec \bar{\theta} d\bar{\theta} 
+ \epsilon \int_0^{\pi/2} e^{-\xi} F_2(\xi) \sec \bar{\theta} d\bar{\theta}   \right] \\
 = 4n_0B_0 \sqrt{\epsilon \beta}R 
\left[S_a + \epsilon S_b \right] \\
 = 4n_0B_0 \sqrt{\epsilon \beta}R S_a
\left[1 + \epsilon \frac{S_b}{S_a} \right]
\end{eqnarray}
Equations (29) and (30) define the integrals $S_a\left( \frac{R}{a} \right)$ and  $S_a\left( \frac{R}{a} \right)$. An appealing feature of equation (31) is that the leading multiplicative term on the right hand side is the value of $\oint n \vec{B} \cdot \vec{ds}$ for the case of uniform density $\bar{n}$. The second term in brackets in (31) is therefore the correction due to the change in density in the Z-Pinch. It is directly proportional to the density modulation index $\epsilon$.  

The parameter $\Gamma \equiv \frac{\epsilon S_b}{S_a}$ gives the fractional error in approximating $\oint n \vec{B} \cdot \vec{ds}$ by $\bar{n} \oint \vec{B} \cdot \vec{ds}$.  Obviously, this is equal to the fractional error in the current inferred from the radioastronomical technique which is the subject of this paper.  
Given the definition of the density modulation index $\epsilon$, one would expect it to be less than unity, say $\epsilon \leq 0.50$, although this is not strictly required in the definition of equation (20).  In any case, it is a parameter which might be observationally accessible. Given an adopted value of $\epsilon$, the correction factor which relates the true current in a plasma to that retrieved by the technique of Section 2 is then dependent on the ratio $\frac{S_b}{S_a}$.  
The integrals   $S_a$ and $S_b$ were evaluated numerically with a {\em Mathematica}
notebook.  The results for a range of values of $\frac{R}{a}$ are given in Table 2. 
\begin{deluxetable}{crrr}
\tablecaption{Validity of Constant Density Approximation\label{tbl-2}}
\tablewidth{0pt}
\tablehead{\colhead{$\frac{R}{a}$} & \colhead{$S_a$} & \colhead{$S_b$} & \colhead{$\frac{S_b}{S_a}$} }
\startdata
 0.1 &  1.682 & 0.645 & 0.383\\
 0.2 &  1.585 & 0.542 & 0.342\\
 0.4 &  1.443 & 0.397 & 0.275\\
 0.6 &  1.331 & 0.297 & 0.223\\
 0.8 &  1.237 & 0.224 & 0.181\\
 1.0 &  1.154 & 0.169 & 0.147\\
 1.4 &  1.015 & 0.098 & 0.096\\
 1.8 &  0.902 & 0.057 & 0.063\\
\enddata
\end{deluxetable}  
The conclusions to be drawn from the calculations summarized in Table 2 are clear.  Even for lines of sight which pass deep in the interior of the current channel, the ratio $\frac{S_b}{S_a} < 1$ by a substantial factor.  The fractional error $\Gamma = \frac{\epsilon S_b}{S_a}$ will be smaller still, even for relatively large compression factors of $\epsilon \simeq 0.5$.  For example, if the line of sight passes within $R = 0.1-0.2 a$ of the center of the filament, and the compression factor $\epsilon = 0.50$, the error in the current retrieved by the method of Sections 2 and 3 would be only 17 - 19 \%.  This is completely negligible in the context of this paper, and indicates that the method described here would be quite accurate.  

For lines of sight that pass further from the center of the filament, the fractional error becomes smaller still, approaching zero as $\frac{R}{a} \gg 1$, as indeed it must. 
\subsection{Analysis of Measurements from the MST}
In this subsection, we consider a test of our method with data from the Faraday rotation experiment on the Madison Symmetric Torus (MST) plasma machine \citep{Brower02,Ding03}. This approach has the advantage of using data from a real plasma, in contrast to the mathematical ideal of Section 4.2.  In addition, as a laboratory experiment, it is diagnosed in a way that the solar corona cannot be.  

In some sense, use of the MST data assures corroboration of the technique.  The goal of the experiments was to measure currents flowing in the MST, and those experiments motivated the current investigation.  However, the analysis of \cite{Brower02} and \cite{Ding03} utilized more experimental information than is available to us, and (at least in part) applied more sophisticated methods of processing the data, which benefitted from a-priori information on the structure of the toroidal plasma.  

In this subsection, we apply the same formulas used in Sections 2 and 3 for 3C228 to published data from the MST \citep{Brower02}, and compare the retrieved current with that known to flow in the MST device.   \cite{Brower02} report that a current of 400 kA flows in the MST during their experiments, and that the mean density (which we adopt for $\bar{n}$) is $10^{19} \mbox{ m}^{-3}$.  The wavelength of the polarized laser beam used to probe the plasma was 432 nm.  

We utilize the data shown in Figure 1 of  \cite{Brower02}, which shows  Faraday rotation of about $5^{\circ}$ for the chord at an offset distance of -17 cm (corresponding to the parameter $R$ defined in the previous subsection), and  Faraday rotation of about  $-3^{\circ}.5$ for the chord at $R=+21$ cm.  We thus have differential Faraday rotation of $\Delta \chi \simeq 8^{\circ}.5$, corresponding to a differential rotation measure $\Delta RM \simeq 8.0 \times 10^5$ rad/m$^2$ between two lines of sight separated by 0.38 meters.  Substitution of this value for $\Delta RM$ into equation (5) yields a current of 241 kA.  

This value itself, resulting from the most straightforward application of the methods used in the study of the corona, is in reasonably good agreement with the known value.  It is within a factor of two of the known, measured current in the MST device.  Similar precision in the measurement of coronal currents would be cause for delirium.  However, we can improve agreement between remotely inferred currents in the MST, using equation (5) of this paper, and the known value for the current.  

The laser polarimeter system of \cite{Brower02} and \cite{Ding03} utilized a number of chords with different values of $R$, so the current profile can be retrieved.  In fact, the change in the functional form of the current density during a sawtooth crash on MST was a main result of \cite{Brower02}.  The inferred toroidal current density profile $J_z(r)$ is plotted in Figure 4 of \cite{Brower02}, and shows that the current channel has a greater radial extent than 0.2m.  Accordingly, the estimate of I immediately above must be an underestimate.  To approximately calculate the current within the two outer chords displayed in Figure 1 of \cite{Brower02}, we represent the current density profile  $J_z(r)$ by a Gaussian, 
\begin{equation}
J_z(r) = J(0)e^{- (r/a)^2}
\end{equation} 
We use plot 4(b) of \cite{Brower02} and the profile measured before a sawtooth crash, for which we estimate values of $J(0) = 1.93 \times 10^6$ A/m$^2$, and $a=0.29$ m.  The total current within two lines of sight separated by a distance $2R$ with the center of the current filament midway between the lines of sight is 
\begin{eqnarray}
I(R) = \pi J(0) a^2 Erf(X)
\end{eqnarray}
where $Erf(X)$ is the Error Function of $X \equiv \frac{R}{a}$.  Using the above values of $J(0)$ and $a$, 
 and with $R=0.19$ m, equation (33) gives $I(R=0.19m) = 331 kA$.  This value is, as expected, in better agreement with the value inferred from equation (5), and indicates a 37 \% error between the ``remote sensing'' and ``true'' value, a difference which includes possible systematic errors in the plasma physics experiment, which are obviously beyond the scope of this paper. In any case, the degree of agreement supports the credibility of the technique discussed in this paper.  

The work of Brower and Ding demonstrates that future coronal differential Faraday rotation observations, with measurable $\Delta RM$ on several lines of sight to a background radio source, could measure not only the total current but the profile of coronal current density, as is done in the MST. To summarize the results of this and the preceding subsection, investigation of a simplified analytical model of coronal current filaments and experimental results from a laboratory plasma have suggested that the approximation of equation (3) results in errors in the retrieved current of, at most, tens of percent. Such an error does not detract from the significance of our results for coronal physics.   
\subsection{Prospects for Improved Modeling of Coronal Currents} 
The analysis of Section 4.2 would be improved through the use of realistic models for the compressive current systems which exist in the solar corona.  As in the case of Sections 4.2 and 4.3, the goal will be to determine the error associated with assuming a constant density in the plasma, so that the associated error in the retrieved current can be estimated. 

A promising set of existing calculations is comprised of the recent coronal simulations of \cite{Gudiksen05} and \cite{Peter06}, which are specifically directed to studies of Joule heating of the corona. The calculations presented in those papers are three dimensional and give the vector magnetic field, plasma density, and current density at all locations in the simulation. The Faraday rotation measure along different lines of sight through the simulation could be calculated and the relationship between differential Faraday rotation and current properties investigated.  

The investigations to date by Gudiksen, Peter, and collaborators have concentrated on the corona at much smaller heliocentric distances than probed by our VLA observations.  However, future calculations could presumably be carried out to simulate conditions at heliocentric distances of $5 - 10 R_{\odot}$. Furthermore, even the present calculations which have been described in the literature could be used to examine the general relation between differential Faraday rotation and current filaments  in a simulated environment patterned after the corona. 
      
\section{Implications for Coronal Heating}
Having provided observational estimates of coronal currents in Section 3 of this paper, there arises the obvious question as to whether these currents can provide the requisite heating of the corona via Joule heating.  Unfortunately, it is not straightforward to answer this question.  The volumetric heating rate, which must be the basis of such a discussion, is $\eta J^2$ where $\eta$ is the electrical resistivity of the plasma, and $J$ is the current density.  To obtain the average coronal heating rate in the region probed by our observations, one must estimate the geometry, extent, and number of the current sheets in a given volume of the corona, in order to estimate both $J$ from a measurement of $I$, and also to calculate the filling factor of the volume in which heating is going on.  This obviously requires a model for the current sheets.  Furthermore, and most importantly, an expression for the resistivity $\eta$ must be chosen.  

We carried out a calculation for the average volumetric heating rate, using two models for the nature of the current sheets responsible for the differential Faraday rotation. These calculations envisioned the corona as  illustrated in Figure 1, with thin, long (along the large-scale coronal field) sheets, in which the sign of the current density would vary from one sheet to the next.  One of the two models assumed that there was equal probability of positive and negative current density, as would be expected in a true turbulent situation, and in which the measured differential Faraday rotation is due to a statistical fluctuation in the number of positive and negative sheets within the Amperian Loop.  The second model assumed a statistical preference for one sign of the current density.  

Formulas were derived for effective volumetric heating rates in both cases.  These heating rates depend on assumed properties of the current sheets, such as thickness, lateral extent, extension along the large scale field, etc.  The heating rates also depend crucially on the resistivity.  Plausible estimates could be obtained for all the geometric properties of these current sheets.  For the resistivity, we used the Spitzer resistivity \citep{Gurnett05} which is based on fully-understood principles and may be assumed to be a lower limit to the true coronal resistivity.  The volumetric heating rate for either model was roughly 6 orders of magnitude less than that estimated by \cite{Cranmer05} as necessary to explain heating in the relevant part of the corona. This calculation indicates that either the currents which are reported here are irrelevant for coronal heating, or that they do play a role in coronal heating, but that the effective resistivity of the coronal plasma at heliocentric distances of $5 - 10 R_{\odot}$ is approximately 6 orders of magnitude greater than the Spitzer value.  A document with a detailed description of these calculations is available from the author. 
\section{Summary and Conclusions}
\begin{enumerate}
\item We have pointed out that a type of polarization measurement (differential Faraday Rotation measurements on an extended radio source)  which can be done with the Very Large Array can yield estimates of electrical currents in the solar corona. The technique could, in principle, be done with some other radio telescopes as well.  This technique is an astronomical adaptation of a diagnostic used in fusion plasmas, and described by \cite{Brower02} and \cite{Ding03}.  
\item The technique has been applied to observations of the radio source 3C228 which were made on two days in August 2003, when the radio source was viewed through the corona at ``impact parameters'' of $6.7$ and $5.2 R_{\odot}$.  Detectable differential Faraday rotation was detected in relatively brief ``events'' on both days, yielding estimated currents of $2.5 \times 10^9$ and $2.3 \times 10^8$ Amperes, respectively.  Another interval of high quality data on one of the days yielded an upper limit to the differential Faraday rotation, and a corresponding upper limit to the current of $8 \times 10^8$ Amperes.  
\item The technique as defined in this paper is dependent on an approximation in which the plasma density is assumed constant in the region within the Amperian Loop which makes the dominant contribution to the rotation measure. The validity of this approximation was investigated in two ways in Section 4.  First, the value of $\oint n \vec{B} \cdot \vec{ds}$ was compared with $\bar{n} \oint \vec{B} \cdot \vec{ds}$ for a Z-Pinch model in which the plasma density is modified in the current filament. Second, the formula used in our radioastronomical observations (eq.[5]) was applied to measurements made with an infrared laser polarimeter on the MST plasma machine \citep{Brower02,Ding03}, and used to estimate the current, which could also be measured independently.  The values agreed to 37 \%.  
\item We explored the implications of the measured currents for  Joule heating of the corona.  The magnitude of the estimated, average volumetric heating rate is rendered uncertain by poorly-known properties of the current sheets or filaments, and by the value of the coronal resistivity. Use of the Spitzer resistivity leads to a value for the volumetric heating rate approximately 6 orders of magnitude less than the value indicated by observations. We conclude that either these currents are irrelevant for coronal heating, or that the true resistivity in the corona exceeds the Spitzer value by several orders of magnitude.  Resolution of this matter obviously lies in a better understanding of the resistivity in a collisionless plasma.  
\end{enumerate}
\acknowledgments
This work was supported at the University of Iowa by grants ATM03-11825, and ATM03-54782 from the National Science Foundation. I thank W.X. Ding and D.L. Brower of the University of California, Los Angeles, for interesting and helpful discussion about Faraday rotation diagnostics of the current profile on the MST Reversed Field Pinch device of the University of Wisconsin.  


\begin{thebibliography}{}
\bibitem[Bird and Edenhofer (1990)]{Bird90} Bird, M.K. and Edenhofer, P. 1990, in Physics of the Inner Heliosphere II, R. Schwenn and E. Marsch, ed., (Springer-Verlag:Berlin), p13
\bibitem[Brower et al (2002)]{Brower02} Brower, D.L. et al. 2002, \prl~88, 185005
\bibitem[Chen (1989)]{Chen89} Chen, J. 1989, \apj~338, 453
\bibitem[Chen and Schuck (2007)]{Chen07} Chen, J. and Schuck, J.W. 2007, \solphys (submitted)
\bibitem[Cranmer and Ballegooijen (2005)]{Cranmer05} Cranmer, S.R. and Ballegooijen, A.A. 2005, \apjs~156, 265
\bibitem[Ding et al (2003)]{Ding03} Ding, W.X. et al. 2003, \prl~90, 035002
\bibitem[Gudiksen and Nordlund (2005)]{Gudiksen05} Gudiksen, B.V. and Nordlund, \AA. 2005, \apj~618, 1020; {\em Erratum: }\apj~623,600
\bibitem[Gurnett and Bhattacharjee (2005)]{Gurnett05} Gurnett, D.A. and Bhattacharjee, A. 2005, Introduction to Plasma Physics, Cambridge University Press
\bibitem[Hollweg et al (1982)]{Hollweg82} Hollweg, J.V., Bird, M.K., Volland, H., Edenhofer, P., Stelzried, C.T., and Seidel, B.L. 1982, \jgr~87, 1
\bibitem[Ingleby et al (2007)]{Ingleby07} Ingleby, L.D., Spangler, S.R., and Whiting, C.A. 2007, \apj (in press, Oct. 10, 2007; astro-ph/0701538)
\bibitem[Krall and Trivelpiece (1973)]{Krall73} Krall, N.A. and Trivelpiece, A.W. 1973, Principles of Plasma Physics, McGraw-Hill
\bibitem[Kraus (1966)]{Kraus66} Kraus, J.D. 1966, Radio Astronomy, McGraw-Hill, p144
\bibitem[Mancuso and Spangler (1999)]{Mancuso99} Mancuso, S. and Spangler, S.R. 1999, \apj~525, 195
\bibitem[Mancuso and Spangler (2000)]{Mancuso00} Mancuso, S. and Spangler, S.R. 2000, \apj~539,480
\bibitem[Parker (1972)]{Parker72} Parker, E.N. 1972, \apj~174,499
\bibitem[Peter et al (2006)]{Peter06} Peter, H., Gudiksen, B.V., and Nordlund, \AA . 2006, \apj~638, 773
\bibitem[Prager (1999)]{Prager99} Prager, S.C. 1999, Plasma Phys. Control. Fusion~41, A129
\bibitem[Sakurai and Spangler (1994)]{Sakurai94} Sakurai, T. and Spangler, S.R. 1994, \apj~434, 773 
\bibitem[Spangler (1999)]{Spangler99} Spangler, S.R. 1999, \apj~522,879
\bibitem[Spangler (2005)]{Spangler05} Spangler, S.R. 2005, \ssr~121,189
\bibitem[Spangler et al (2007)]{Spangler07} Spangler, S.R., Spitler, L.G., Miralles, M.P., Cranmer, S.R., Raymond, J.C., and Kasper, J.C.  2007, (in preparation)
\end{thebibliography}
\end{document}